\documentclass[]{JHEP3}
\usepackage{cite,graphicx,amsmath,amssymb,xspace}

\newcommand{\be}{\begin{equation}}
\newcommand{\ee}{\end{equation}}
\newcommand{\beq}{\begin{eqnarray}}
\newcommand{\eeq}{\end{eqnarray}}

\newcommand{\gone}[1]{{}}

\newcommand{\Hspace}{\ensuremath{H}\xspace}
\newcommand{\fspace}{\ensuremath{f}\xspace}
\newcommand{\Qspace}{\ensuremath{Q}\xspace}
\newcommand{\Rspace}{\ensuremath{R}\xspace}

\preprint{hep-th/0607135}

\title{Effective descriptions of branes on non-geometric tori}

\author{Ian Ellwood and Akikazu Hashimoto\\
Department of Physics\\
University of Wisconsin, Madison, WI 53706\\
E-mail: \email{iellwood@physics.wisc.edu, aki@physics.wisc.edu}}

\abstract{We investigate the low-energy effective description of non-geometric
compactifications constructed by T-dualizing two or three of the
directions of a $T^3$ with non-vanishing $H$-flux. Our approach is to
introduce a D3-brane in these geometries and to take an appropriate
decoupling limit. In the case of two T-dualities, we find at low
energies a non-commutative $T^2$ fibered non-trivially over an
$S^1$. In the UV this theory is still decoupled from gravity, but is
dual to a little string theory with flavor.  For the case of three
T-dualities, we do not find a sensible decoupling limit, casting doubt
on this geometry as a low-energy effective notion in critical
string theory.  However, by studying a topological toy model in this
background, we find a non-associative geometry similar to one found by
Bouwknegt, Hannabuss, and Mathai.
}


\begin{document}

\section{Introduction}

Compactifications on non-geometric spaces have emerged as a new
potential class of string theory vacua
\cite{Hellerman:2002ax,Dabholkar:2002sy,Kachru:2002sk}.  The
non-geometric nature of these spaces arises because some of the
transformations that glue the patches together include
U-dualities as well as the standard diffeomorphisms.  A useful
prototype non-geometric space is found by T-dualizing a three
dimensional torus with a non-vanishing $H$-flux: T-dualizing on one
cycle gives rise to a Scherk-Schwarz twisted torus
\cite{Scherk:1979zr,Alvarez:1993qi,Hull:1998vy,Gurrieri:2002wz,Bouwknegt:2003vb}  which is purely geometric. However, T-dualizing
on two cycles gives rise to a space in which one of the cycles is
periodic up to T-duality, which mixes momentum and winding modes \cite{Kachru:2002sk, Lowe:2003qy, Mathai:2004qq,Mathai:2004qc,Mathai:2005fd,Shelton:2005cf,Dabholkar:2005ve, Benmachiche:2006df, Shelton:2006fd, Hull:2006va, Hull:2006qs}.  As
such, in this space, geometric notions such as the metric and the
background $B$-field are only well-defined locally.

After T-dualizing twice, one can contemplate T-dualizing along the
third direction of the $T^3$. Naively, a $T^3$ with uniform $H$-flux
is isometric under shifts in three independent dimensions. However,
the Buscher rules for T-duality require that the two-form potential
$B$ be uniform\cite{Buscher:1987qj,Giveon:1994fu}, and in order for
the three form field strength $H$ to be uniform, the two form
potential must break at least one of the translation isometries.
Thus, the standard Buscher rules do not apply.

Recently, Shelton, Taylor, and Wecht proposed an interpretation of the
third T-dual of a torus with $H$-flux as an example of a consistent
non-geometric compactification \cite{Shelton:2005cf}. They also
introduced a nomenclature which we follow: the \Hspace,
\fspace, \Qspace, and \Rspace-spaces correspond to a $T^3$ with
$H$-flux T-dualized zero, one, two, and three times, respectively. The
\Qspace-space is perhaps the simplest example of a non-geometric
compactification that mixes momentum and winding modes. If the
interpretation of \cite{Shelton:2005cf} is correct, the \Rspace-space
is an example of a space that is more non-geometric than the
\Qspace-space.\footnote{These fluxes are also interesting from the
``cosmological billiards'' perspective, since they correspond to
interesting roots of $E_{10}$ \cite{Damour:2002et,Brown:2004jb}. We thank Ori Ganor
for bring this point to our attention.}

Space-time geometry in string theory is an approximate notion which is
valid only when the scale of all the geometric features are much
larger than string scale. When the size of a compact manifold becomes
comparable to the string scale, geometric notions break down due to the
non-locality intrinsic to the fact that a string is an extended
object. In order to isolate the novel geometric features from the
generic non-locality effects of string theory, one takes the
decoupling limit, $\alpha'\to 0$, keeping the size of the compact manifold
finite. Geometric notions acquire precise meaning in this
limit. However, not all space-time geometries one obtains in this way
are ordinary geometric spaces.  More exotic spaces, such as the
non-commutative plane, can arise as decoupling limits of string
theory.

One can investigate the geometric features of the \Qspace-space and
the \Rspace-space along similar lines.  Our strategy is to use the
properties of field theories defined on these spaces as a probe of the
geometry.  In string theory, this can be implemented by introducing a
D3-brane filling the space and taking the decoupling
limit.\footnote{Adding D-branes to various T-duals of $\Hspace$-space
was also studied in \cite{Hull:2004in,Lawrence:2006ma,Marchesano:2006ns}.} This
gives rise to a non-trivial effective dynamics of open strings ending on
a D-brane in a presence of an NSNS $B$-field background. 

Because of the presence of the $B$-field, it is natural to expect some connection between the \Qspace/\Rspace-spaces and
non-commutative geometry.  Indeed, in the case of the
\Qspace-space, we find a familiar non-commutative theory whose
UV completion is a little string theory coupled with flavor. On the
other hand, we do not find a clean decoupling limit for a
theory defined on \Rspace-space. The
\Rspace-space does not appear to admit an effective description as a
smooth macroscopic structure decoupled from gravity.

This article is organized as follows. In section 2 we review the
supergravity background giving rise to the \Hspace, \fspace,
and \Qspace  spaces.  In section 3, we describe the decoupled
theory of D3-branes in the \Qspace-space and its UV completion. In section
4, we comment on the status of \Rspace-space.  In section 5 we discuss
a toy model for \Rspace-space physics.  We end with some concluding
remarks in section 6.

\section{Torus with $H$-flux and its T-duals}

In this section, we review the spaces \Hspace, \fspace,
\Qspace, and \Rspace, in the nomenclature of \cite{Shelton:2005cf}. We
will also review the warped supergravity background of smeared
NS5-branes that properly takes into account the gravitational
backreaction of the non-vanishing $H$-flux.

\subsection{The \Hspace-space\label{hspacesec}}

Consider a $T^3$ with $H$-flux whose coordinates are given by $x_i$
with periodicities $L_i$ for $i = 1,2,3$. When written explicitly, the
background $B$-field breaks at least one of the infinitesimal
translation symmetries:
\be 
  ds^2 = dx_1^2+ dx_2^2+ dx_3^2, 
\qquad 
  B = b\, x_1\,  dx_2 \wedge dx_3, 
\qquad 
  H = dB, \qquad b={(2 \pi)^2 N \alpha'\over L_1 L_2 L_3} 
\label{hspace} \ . \ee
The $H$-flux is quantized so that $N$ is an integer. We refer to this space as the \Hspace-space using
a notation similar to
\cite{Shelton:2005cf}.

\subsection{The \fspace-space}

T-dualizing along $x_3$ gives rise to a purely metrical background
\be ds^2 = dx_1^2 + dx_2^2 + (d \tilde x_3 + b x_1 d x_2)^2   \label{fspace},\ee
with twisted boundary conditions,
\be x_1 \sim x_1 + n_1 L_1 , \qquad x_2 \sim x_2 + n_2 L_2 , \qquad \tilde x_3 = \tilde x_3 + n_3 \tilde L_3 -   n_1  b L_1 x_2  , \qquad \tilde L_3 = {(2 \pi)^2 \alpha' \over L_3}. \ee
This space is also known as the {\it nil manifold} or the {\it twisted
torus} and is topologically distinct from the ordinary
torus; for example, it has $H_1({\bf Z})$ given by ${\bf Z} \times {\bf Z}
\times {\bf Z}_N$ \cite{Kachru:2002sk,Marchesano:2006ns}. Following
\cite{Shelton:2005cf} we refer to it as the
\fspace-space.

Introducing dimensionless coordinates,
\be x_i = L_i y_i, \ee 
the metric and the boundary condition becomes more transparent: 
\be ds^2 = L_1^2 dy_1^2 + L_2^2 dy^2 + L_3^2 (d\tilde y_3 + N y_1 dy_2) ,\ee
\be y_1 \sim y_1 + n_1 , \qquad y_2 \sim y_2 + n_2 , \qquad \tilde y_3 = \tilde y_3 + n_3  -   n_1  N y_2   \ . \ee
In this form, it is also apparent that $N$ is quantized to be an integer.

\subsection{The \Qspace-space}

Further T-dualizing along the $x_2$ direction gives the background,
\beq ds^2 &=& dx_1^2 + {1 \over 1+ b^2 x_1^2} (d \tilde x_2^2 + d \tilde x_3^2) ,\cr
B & = & {b x_1 \over 1 + b^2 x_1^2}  d\tilde x_2 \wedge d \tilde x_3,  \label{qspace} \\
e^{\phi-\phi_0} & = &  {1 \over \sqrt{1 + b^2 x_1^2}}\nonumber \ .  \eeq
This space is called the \Qspace-space in \cite{Shelton:2005cf}.  This
background is periodic in the $x_1$ direction up to a T-duality  of
the $\tilde x_2$-$\tilde x_3$ torus, which exchanges momentum and
winding modes. As is clear from (\ref{qspace}), the metric and
$B$-field are locally defined, but globally are not manifestly
periodic in the $x_1$ direction.  As such we take it as a prototypical
example of a ``non-geometric'' space
\cite{Hellerman:2002ax,Dabholkar:2002sy,Kachru:2002sk}.

\subsection{The \Rspace-space}

It is not completely clear that the \Qspace-space can be further
T-dualized along the $x_1$ coordinate.  Translation along the $x_1$
direction is, after all, not (even locally) an isometry. Assuming that a
T-dual does exist, this space was named the \Rspace-space by
\cite{Shelton:2005cf}.

\subsection{Smeared NS5-brane background}

As is, the $T^3$ with non-vanishing $H$-flux described in section
\ref{hspacesec} is not a consistent closed string background since it
does not take into account the gravitational backreaction of the
$H$-flux. One convenient way to build a consistent background is to
start with the NS5-brane background (extended along the 56789
directions),
\beq
ds^2 & = & -dt^2 + dx_5^2 + \ldots +  dx_9^2 + f(r) (dx_1^2 + dx_2^2 + dx_3^2 + dx_4^2), \cr
H & = &  *(dt \wedge dx_5 \wedge ... dx_9 \wedge d f^{-1}), \cr
e^\phi & = & g_s f(r)^{1/2}, \cr
f(r) &=& 1 + {m \alpha' \over r^2}, \label{ns5bg}
\eeq
which is a magnetic source of the 3-form $H$-field, and to smear it
along three of the four transverse coordinates so that the
supergravity background becomes \cite{Hull:1998vy, Gurrieri:2002wz, Lowe:2003qy}
\beq
ds^2 & = & -dt^2 + dx_5^2 + \ldots +  dx_9^2 + f(r) (dx_1^2 + dx_2^2 + dx_3^2 + dz^2), \cr
H & = & dB, \cr
B & = &  {(2 \pi)^2 N \alpha' \over L_1 L_2 L_3} (x_1-x_1^0) \,  dx_2 \wedge dx_3,\cr
e^\phi & = & g_s f(z)^{1/2}, \cr
f(z) &=& f_0 - {(2 \pi)^2 N\alpha' (|z|+z)\over 2 L_1 L_2 L_3} \label{background} \ . 
\eeq
The smeared NS5-branes are located at $z=0$. The parameter $f_0$,
which we take to be positive, is otherwise freely adjustable. The
parameter $x_1^0$ is just an additive constant for the $x_1$
coordinate. We have tuned the charges at infinity so that $H=0$ for
$z<0$. The smeared NS5-brane acts as a domain wall source for a
uniform three-form field strength $H$ in the region $z>0$.  See
figure \ref{figaa} for an illustration.

\FIGURE{
\centerline{
\begin{picture}(254,164)(0,0)
\put(106,145){$f(z)$}
\put(37,117){${1 \over (2 \pi)^2 \alpha'}\int H=0$}
\put(123,117){${1 \over (2 \pi)^2 \alpha'}\int H=N$}
\put(229,44){$z$}
\includegraphics{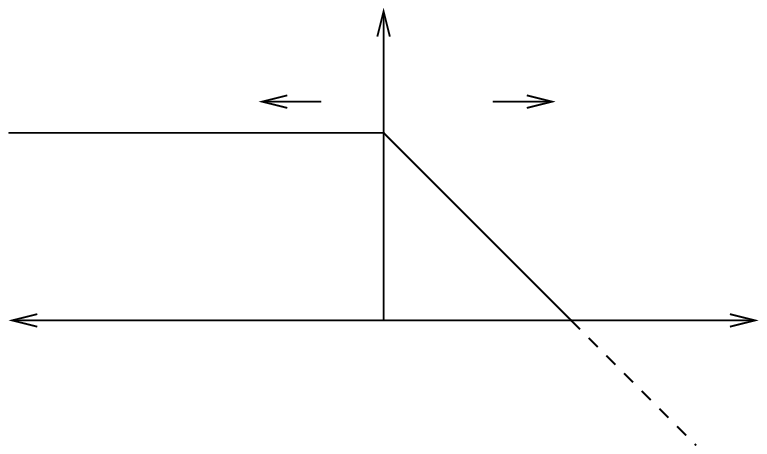}
\end{picture}
}
\caption{\footnotesize{Profile of the harmonic function $f(z)$. \label{figaa}}}
}

This is not the only way to construct a solution to the supergravity
equations of motion with non-vanishing $H$-flux threading a $T^3$.
This solution, however, is convenient in that it preserves 16 of the
32 supersymmetries in type IIA or type IIB supergravity. We will
consider T-duals of (\ref{background}) in order to consistently embed
the \Hspace, \fspace, \Qspace, and \Rspace spaces into type IIA/B
supergravity.

\section{Decoupled theory of D3-branes in $Q$-space\label{qsec}}

In this section, we derive an effective geometry for D3-branes wrapping the \Qspace-space. This setup
is equivalent to starting with a D1-brane wrapping the $x_1$ direction of the $H$-space and T-dualizing along the $x_2$ and
$x_3$ directions\footnote{We use a $\bullet$ to denote directions along which a brane is extended and $\equiv$ to denote directions along which it is smeared.};\\
\be \begin{tabular}{cccccccccccc}
&0&1&2&3&4&5&6&7&8&9 \cr
NS5 & $\bullet$ & $\equiv$&$\equiv$&$\equiv$& & $\bullet$& $\bullet$ & $\bullet$ & $\bullet$ & $\bullet$  \cr
D1 & $\bullet$ &  $\bullet$  &  & &
\end{tabular} \label{qconf} \ee
This configuration preserves 8 of the 32 supersymmetries of type
IIB, as can be seen by T-dualizing along the $x_5$ and the $x_6$
coordinates, S-dualizing, and counting the number of relatively
transverse coordinates of the branes.

Since the $H$-space metric (\ref{background}) is manifestly isometric along the $x_2$ and
$x_3$ directions, it is straightforward to T-dualize along these
coordinates, giving the background,
\beq ds^2 & = & f(U) dx_1^2 + \frac{f(U)}{ f(U)^2 + \left({N \tilde L_2 \tilde L_3 \over \alpha' L_1}(x_1-x_1^0)\right)^2 } (d \tilde x_2^2 + d \tilde x_3^2), \cr
B_{23} & = & 
\frac{{N \tilde L_2 \tilde L_3 \over \alpha' L_1}(x_1 - x_1^0)}{ f(U)^2 + \left({N \tilde L_2 \tilde L_3 \over \alpha' L_1}(x_1-x_1^0)\right)^2 }, \cr
f(U) &  = & f_0 - \frac{N \tilde L_2 \tilde L_3 (|U| +U ) }{ 2 L_1}, \label{Qsugra}\eeq
where
\be \tilde L_{2,3} = (2 \pi)^2 {\alpha' \over L_{2,3}} \ee
is the period of the dual coordinates $\tilde x_{2,3}$. The transverse coordinate $x_4$ has been scaled as
\be x_4 = \alpha' U\  \ee
so that $U$ parameterizes the vacuum expectation value of a scalar
field polarized along the $x_4$ direction.

In order to take the decoupling limit, we send $\alpha'
\rightarrow 0$ keeping the field theory parameters $L_1$, $\tilde
L_{2,3}$ and $U$ fixed. In this limit, the metric and the $B$-field degenerate; however, since we are interested in the effective
dynamics of decoupled open strings, we should not be concerned about
the scaling of the closed string variables $g$ and $B$.  Instead, we
should consider the open string metric and non-commutativity parameter
along the lines of
\cite{Seiberg:1999vs,Hashimoto:2002nr,Dolan:2002px,Hashimoto:2004pb,Hashimoto:2005hy};
\be (g + B)^{-1} = G + \frac{\theta }{ 2 \pi \alpha'}. \ee
Explicitly, we find
\be G^{ij} = f(U) \left(\begin{array}{ccc}1 & & \\& 1 & \\ &&1 \end{array}\right)^{ij} , \qquad 
\Theta^{23} = {2 \pi \theta^{23} \over \tilde L_2 \tilde L_3} =  N \left({x_1-x_1^0  \over L_1}\right)\ . \label{openparam} \ee
The dimensionless non-commutativity parameter $\Theta^{23}$ is simply
the non-commutativity parameter $2 \pi \theta^{23}$ divided by the
volume of the torus $\tilde L_2 \tilde L_3$.  Rather remarkably, both
the open string metric and the non-commutativity parameters remain
finite in the limit.

The manifestation of the non-geometric character of the \Qspace-space
in the decoupled theory is now transparent. Since the
non-commutativity parameter $\Theta^{23}$ depends explicitly on $x_1$,
one cannot naively make this coordinate periodic. However, under a
discrete shift $x_1 \rightarrow x_1 + L_1$, the dimensionless
non-commutativity parameter $\Theta^{23}$ shifts by $N$. Such a shift
of $\Theta^{23}$ by an integer is an example of Morita equivalence,
which acts as an $SL(2,Z)$ transformation on the parameters of
non-commutative torus (using the notation explained in the appendix of
\cite{Hashimoto:1999yj}) as follows:
\begin{eqnarray}
&&\tilde \Theta = {c + d \Theta \over a + b \Theta}, \qquad 
\tilde \Phi = (a + b \Theta)^2 \Phi - b ( a + b \Theta), \qquad 
\tilde{\Sigma} = (a + b \Theta) \Sigma, \nonumber \\
&&\tilde{g}_{YM}^2 = (a + b \Theta) g_{YM}^2, \qquad 
\left(\begin{array}{c} \tilde m \\ \tilde N \end{array}\right)
= \left(\begin{array}{cc} a & b \\ c & d \end{array}\right)
\left( \begin{array}{c} m \\ N \end{array}\right) \ . 
\end{eqnarray}
Therefore, in the \Qspace-space, the $x_1$ coordinate is only periodic
up to a Morita transformation. Morita equivalence is precisely the
structure inherited from T-duality in the decoupling limit.  It is
therefore natural to find a compactification that
identifies shifts in $x_1$ via Morita equivalence emerging as a
decoupling limit of a compactification that identifies shifts in $x_1$
via T-duality.

This geometrical structure, which can be viewed as a field of
non-commutative tori fibered over a circle, also appears in the works
of
\cite{Mathai:2004qq,Mathai:2004qc,Mathai:2005fd,Bouwknegt:2004ap}. In
our work, we emphasize the fact that this structure has a physical
origin as the decoupled theory of the open string excitations living
on the world volume of a D3-brane embedded into the $Q$-space.

Although the motivation was somewhat different, most of the features
of the decoupled field theory on D3-branes in the \Qspace-space were
first worked out in \cite{Lowe:2003qy}. One feature, which did not
get emphasized in \cite{Lowe:2003qy}, however, is that the
open-string metric (\ref{openparam}) is warped in the transverse
coordinate $U$. In fact, there will be a singularity at some finite
value of $U$. Since a typical string fluctuates by a size of the order
of $l_s$, which is much larger than the distance to the singularity,
which is order $l_s^2$, the open string dynamics is strongly
influenced by the presence of the singularity.  Hence, it would be
premature to assume, for example, that the low-energy effective theory
is precisely ${\cal N}=4$ supersymmetric Yang-Mills theory
non-commutatized by the position dependent $\Theta$ given in
(\ref{openparam}). Such a theory would have an unbroken $SO(6)$
R-symmetry group, which is clearly broken by the warping along the $U$
direction.

Since there is no physical scale other than the scale of
compactification and the non-commutativity, one can also think of this
system as having non-commutativity parameters that are different for
different $U(1)$ subsectors when the gauge group is broken by turning
on the vacuum expectation values for the transverse scalar field along
the warped coordinate $U$. Models with these features have been
considered before \cite{Dolan:2000mz,Dasgupta:2000um}.  The authors of
\cite{Dasgupta:2000um} referred to these models as ``non-abelian
geometry.''

In order to infer the correct low-energy effective action for the
decoupled theory, one must analyze the dynamics of open strings in
this background in some detail.\footnote{The authors of
\cite{Dasgupta:2000um} proposed a generic non-abelian $*$-product, but
we do not see how such a product properly incorporates the dynamics of
open strings in this background.} This seems like a serious technical
challenge in light of the fact that the world sheet sigma model for
strings propagating in the background of smeared NS5-branes does not
appear to be exactly solvable.\footnote{In order to study explicit
realization of non-abelian geometry, one can instead consider simpler
construction based on Melvin universes which are solvable
\cite{melvinNA}.}

\subsection{UV completion of the \Qspace-space effective geometry}

That there are singularities at a finite distance in moduli-space
suggests that the effective description based on non-commutative field
theory is breaking down because of certain states that were
integrated out.  In the remainder of this section, we will work out a
UV completion of this theory that resolves the singularity while
keeping gravity decoupled.

The singularity of the smeared NS5-brane background is closely related
to the singularity in type I' theory that one encounters in the
heterotic/type I duality \cite{Polchinski:1995df}. The mechanism that
resolves the singularity is also similar. To see this more explicitly,
it is useful to embed the 1+1 dimensional effective dynamics of
D1-branes in the configuration (\ref{qconf}) as a dimensional
reduction of 3+1 dimensional system oriented as follows:

\noindent {\it Step I:}
\be
\begin{tabular}{cccccccccccc}
&0&1&2&3&4&5&6&7&8&9 \cr
NS5 & $\bullet$ & $\equiv$&$\equiv$&$\equiv$& & $\bullet$& $\bullet$ & $\bullet$ & $\bullet$ & $\bullet$\cr
D3 & $\bullet$ &$\bullet$ &&&  & $\bullet$ & $\bullet$  \end{tabular}
\ee
The coordinates $x_5$ and $x_6$ are taken to be compact with
period $L_{5,6}$, which will remain finite in the scaling
limit. ($L_{5,6}$ can be taken to be small compared to other scales of
the problem at the very end.) In order for this background to
yield the
\Qspace-space after two T-dualities, we scale the parameters of the
compactification as follows:\footnote{Factors
of $2$ and $\pi$ are left out to prevent cluttering.}

\be g_I = g, \qquad \alpha'_I = \alpha', \qquad 
 {L_1}\raisebox{-0.65ex}{\rule{0ex}{1ex}}_{I} = L_1, 
\qquad {L_{2,3}}_I = {\alpha' \over \tilde L_{23}}.\ee

\noindent {\it Step II: $ST_{123}$-duality}

Next, we S-dualize and then perform T-duality along the $x_1$, $x_2$, and $x_3$ directions. This gives the configuration,
\be
\begin{tabular}{cccccccccccc}
&0&1&2&3&4&5&6&7&8&9 \cr
D8 & $\bullet$ & $\bullet$ &$\bullet$&$\bullet$& & $\bullet$& $\bullet$ & $\bullet$ & $\bullet$ & $\bullet$\cr
D4 & $\bullet$ & &$\bullet$&$\bullet$&  & $\bullet$ & $\bullet$  \end{tabular}\ee
where the parameters after the duality scale as
\be g_{II} = {g L_2 L_3 \over L_1 {\alpha'_{II}}^{1/2}}, \qquad \alpha'_{II} = g \alpha', \qquad 
{L_1}\raisebox{-0.65ex}{\rule{0ex}{1ex}}_{II} = {g \alpha' \over L_1}, \qquad {L_{2,3}}_{II} = g \tilde L_{2,3}\ . \ee
This is the same brane configuration discussed in
\cite{Seiberg:1996bd,Douglas:1996xp,Morrison:1996xf} that gives rise
to non-trivial fixed points in 4+1 dimensions precisely when the
D4-brane is placed at the singularity. The singularity in this context
arises from integrating out the strings stretching between the
D4-brane and the D8-brane. It is convenient to view this system as a
decompactification limit of a type I' theory given by separating the
D8-branes from the O8-branes. To match the harmonic function profile
with what is illustrated in figure \ref{figaa}, one should imagine
eight D8-branes and an O8 brane to the far left, N D8-branes at $z=0$,
and $(8-N)$ D8-branes and an O8-brane to the far right.

\noindent {\it Step III: 1-11 Flip}

The only thing which makes this description unsuitable as an effective
description of the \Qspace-space is the small size of the period of the $x_1$ direction. This can be rectified by performing a 1-11
flip, which yields the configuration,
\be \begin{tabular}{cccccccccccc}
&0&1&2&3&4&5&6&7&8&9 \cr
D8 & $\bullet$ & $\bullet$ &$\bullet$&$\bullet$& & $\bullet$& $\bullet$ & $\bullet$ & $\bullet$ & $\bullet$\cr
NS5 & $\bullet$ & $\bullet$ &$\bullet$&$\bullet$&  & $\bullet$ & $\bullet$  \end{tabular} \label{lst-flavor}\ee
with parameters
\be g_{III}  = {\alpha' \sqrt{g} \over L_1 \sqrt{\tilde L_2 \tilde L_3} }, \qquad  
{L_1}\raisebox{-0.65ex}{\rule{0ex}{1ex}}_{III} = {g \tilde L_2 \tilde L_3 \over L_1}, \qquad 
\alpha'_{III}= g \tilde L_2 \tilde L_3
, \qquad {L_{2,3}}_{III} = g \tilde L_{2,3}\ . \ee
Notice that as $\alpha'\rightarrow 0$, $g_{III}$ goes to zero as
well, while $\alpha'_{III}$ and the various length scales associated
with the world volume of the NS5-brane remain fixed. This is the
standard decoupling limit of little string theory.

An important set of light degrees of freedom in this limit are the D2-branes
stretching between the NS5-brane and the D8-branes.  In the limit, the
D2-brane behaves effectively as a string. If the distance separating
the NS5-brane and the D8/O8-brane is of the order $\alpha' U$, then
the tension of this effective string is
\be T = {1 \over g_{III} {\alpha'_{III}}^{3/2} } \alpha' U = 
 {L_1 U \over g^2 \tilde L_2 \tilde L_3},\ee
which remains finite as $\alpha' \to 0$. These states are the
analogues of ``fundamental matter'' for little string theory, along
the lines discussed for the case of D3-branes in
\cite{Aharony:1998xz}. It is also the IIA reduction (along one of the
dimensions transverse to the M5-brane) of the non-critical string
theory introduced in \cite{Ganor:1996mu}. The conclusion is that gauge
theory on \Qspace-space is a low-energy effective description of
``little string theory with flavor,'' which is decoupled from gravity.

\section{Low-energy effective description of the $R$-space\label{sec3}}

We will now describe what happens if one attempts to repeat the story
for the \Rspace space.  That we do not have an explicit supergravity
solution describing the \Rspace-space analogous to (\ref{Qsugra}) prevents us
from directly inserting a D3-brane as we did in the previous section.
What we can do instead is to insert a D0-brane in the \Hspace-space,
and study the low-energy effective dynamics, while scaling the size of
the torus as $L_{1,2,3} \sim \alpha' / \tilde L_{1,2,3}$. This scaling
isolates the winding modes while decoupling the momentum modes as
$\alpha' \rightarrow 0$. Re-interpreting the winding modes of one
geometry as Kaluza-Klein excitations of some other geometry
essentially amounts to performing a T-duality.\footnote{One can in
fact think of our construction as the generalization of
\cite{Taylor:1996ik,Ganor:1996zk} on $T^3$ with non-vanishing
$H$-field.} In fact, had the $B$-field been taken to be a constant,
this is precisely the approach taken in
\cite{Connes:1997cr,Douglas:1997fm,Cheung:1998nr} to construct
ordinary non-commutative spaces as a decoupling limit.

One can easily check that D0-branes in the background of smeared
NS5-branes,
\be \begin{tabular}{cccccccccccc}
&0&1&2&3&4&5&6&7&8&9 \cr
NS5 & $\bullet$ & $\equiv$&$\equiv$&$\equiv$& & $\bullet$& $\bullet$ & $\bullet$ & $\bullet$ & $\bullet$\cr
D0 & $\bullet$ &   \end{tabular} \label{D0inR}\ee
break all supersymmetries and, hence, there will be a potential for the
D0-brane to roll toward the NS5-branes.  An alternative setup is to consider
a D1-brane extended along the warped direction:
\be \begin{tabular}{cccccccccccc}
&0&1&2&3&4&5&6&7&8&9 \cr
NS5 & $\bullet$ & $\equiv$&$\equiv$&$\equiv$& & $\bullet$& $\bullet$ & $\bullet$ & $\bullet$ & $\bullet$\cr
D1 & $\bullet$ &&&& $\bullet$   \end{tabular} \label{Rbraneconf}\ee
This configuration is supersymmetric and static.  In order to isolate
the low-energy effective dynamics on the \Rspace space, we scale our parameters as
\be g_s = {g_{YM4}^2 \alpha' \over \tilde L_1 \tilde L_2 \tilde L_3} , \qquad L_{123}= {\alpha' \over \tilde L_{1,2,3}}, \ee
so that the volume $\tilde L_1 \tilde L_2 \tilde L_3$ and the gauge coupling
$g_{YM4}^2$ of the 4+1 dimensional Yang-Mills theory are finite  after T-dualizing
along $x_1$, $x_2$, and $x_3$ coordinates.

There is a problem with interpreting such a construction as
effectively giving rise to the \Rspace-space. The smeared NS5-brane
background is still given by (\ref{background}), but now the
harmonic function $f(z)$, when written in terms of $\alpha'$ and the
variables that are kept finite in the scaling limit, takes the form
\be f(z) = f_0 - { N  \tilde L_1 \tilde L_2 \tilde L_3 (|z|+z)\over 2 (2 \pi)^4 \alpha'^2} \ . \label{Rfz} \ee
For the \Rspace-space to be a useful effective notion, one would like $f(z)$ to be of order 1 for a sufficiently wide range
of values of $z$. Clearly, this is not the case in the limit $\alpha'
\rightarrow 0$.

Such severe warping appears to make the \Rspace-space problematic as a
low-energy effective notion.  The appearance of a strong gravitational
backreaction is to be expected since we start by shrinking the volume
of the torus in the \Hspace-space, where the total flux is kept
fixed. As we shrink the torus, the energy density associated with the
flux increases, giving rise to stronger backreaction on the geometry.
Such a strong gravitational backreaction is also potentially
problematic for the effective description of the \Qspace-space
considered in the previous section. It is the combination of the fact
that the probe branes can be arranged to be localized in the
warped direction, and that the amount of squeezing of the flux is
milder in the scaling relevant to the \Qspace-space that allows for a
smooth decoupling limit.

In spite of the strong warping, one could contemplate performing a duality transformation on (\ref{Rbraneconf}) to attempt to identify the analogue of (\ref{lst-flavor}) for the \Rspace-space case.  

\noindent {\it Step I: $T_{56}$ duality:}
\be \begin{tabular}{cccccccccccc}
&0&1&2&3&4&5&6&7&8&9 \cr
NS5 & $\bullet$ & $\equiv$&$\equiv$&$\equiv$& & $\bullet$& $\bullet$ & $\bullet$ & $\bullet$ & $\bullet$\cr
D3 & $\bullet$ &&&& $\bullet$ & $\bullet$ & $\bullet$  \end{tabular} \ee
\be g_I = {g_s L_5 L_6\over \alpha'} = {g_{YM4}^2 L_5 L_6 \over \tilde L_1 \tilde L_2 \tilde L_3} = \mbox{finite}, \qquad {L_{1,2,3}}_I = {\alpha' \over \tilde L_{1,2,3}}, \qquad
{L_{5,6}}_I = \mbox{finite}\ . \label{RconfII}\ee

\noindent {\it Step II: $ST_{123}$ duality:}

This duality transforms (\ref{RconfII}) to
\be
\begin{tabular}{cccccccccccc}
&0&1&2&3&4&5&6&7&8&9 \cr
D8 & $\bullet$ & $\bullet$ & $\bullet$ & $\bullet$ &  & $\bullet$& $\bullet$ & $\bullet$ & $\bullet$ & $\bullet$\cr
D6 & $\bullet$ &$\bullet$ &$\bullet$ &$\bullet$ & $\bullet$ & $\bullet$ & $\bullet$  \end{tabular}
\ee
with the parameters
\be g_{II} = {
 g_I^2 \tilde L_1 \tilde L_2 \tilde L_3
 \over {\alpha'_{II}}^{3/2}}, \qquad \alpha'_{II} = g_I \alpha'_I, \qquad {L_{1,2,3}}_{II} = g_I {\tilde L_{1,2,3}} \ . 
\ee
This gives rise to a low-energy effective coupling for the D6-brane,
\be g_{YM6}^2 = g_{II} {\alpha'}_{II}^{3/2} =  g_I^2 \tilde L_1 \tilde L_2 \tilde L_3 = \mbox{finite} \ . 
\ee
In fact, this configuration is the infinite volume limit of the 6+1
dimensional gauge theory discussed in \cite{Hanany:1997gh}.  Unlike
the NS5-branes considered in the previous section, D6-branes do not
have a scaling limit that decouples the gauge dynamics from
gravity. Nonetheless, it is tempting to speculate that the $R$-space
emerges as a low-energy effective description of this dynamical
system.  Unfortunately, as the slope of the harmonic function (\ref{Rfz}) diverges
in the $\alpha' \rightarrow 0$ limit, the range of the coordinate $z$ for
which this system is dual to an $R$-space shrinks to zero size, making
such an interpretation doubtful.

Had we considered instead the non-supersymmetric D0-brane
configuration shown in (\ref{D0inR}), the warping would cause the
D0-brane to roll toward the NS5-branes in a time-scale that gets
arbitrarily short as $\alpha'$ is sent to zero.

All of these problems are a consequence of generic gravitational
backreaction effects arising from quantized fluxes in a small volume.
While we demonstrated the difficulty only for the specific case of
$H$-flux generated by the background of smeared NS5-branes, it seems
reasonable to expect that any realization of $H$-flux on a torus would
lead to similar difficulties.\footnote{Recently, a large class of
solutions of the four dimensional effective field theory, where the
effects of the fluxes in compact dimensions are encoded in the
superpotential, were constructed in \cite{Shelton:2006fd}. It would be
interesting to see if any of these constructions would allow a smooth
decoupling limit to be taken along the lines discussed in this paper.
To study this issue, however, it is essential to first find an
explicit lift of these solutions to a solution of supergravity in ten
dimensions similar to  (\ref{background}).}  We are therefore propose
that the
\Rspace-space does not exist as a low-energy effective notion
decoupled from the stringy effects.

\section{Comments on possible connections to the non-associative tori}

One of the main conclusions of this article is the observation that a
sensible decoupling limit of low-energy open strings in
\Rspace-space does not appear to exist.  As we saw in the previous
section, the main cause of this difficulty is the strong gravitational
backreaction in the dual \Hspace-space description of the background
geometry.  It turns out, however, that an intriguing non-associative
geometry, similar to the one described in \cite{Bouwknegt:2004ap},
emerges if one naively ignores the gravitational backreaction. The
fact that we are ignoring the gravitational backreaction, which is
necessary in order to have a consistent background for string
perturbation theory, makes the physical interpretation of this
mathematical structure in terms of string theory less
clear. Nonetheless, a connection to some kind of non-associative
geometry as an effective description of \Rspace-space is sufficiently
intriguing that we felt it worth illustrating. Our
hope is that this discussion can be made more transparent in the
future.

\subsection{Review of dual lattice formulation of non-commutative geometry}

We begin the discussion by recalling the approach used in
\cite{Connes:1997cr,Douglas:1997fm,Cheung:1998nr} to describe ordinary
non-commutative spaces. A D2-brane in a $B$-field background becomes a
non-commutative gauge theory in a certain scaling limit. Instead of
the D2-branes in a $B$-field background, however, one can also
consider D0-branes in the T-dual torus of size $L=\alpha'/\tilde L$,
which also has a non-vanishing $B$-field.

Before T-dualizing, the non-commutativity manifests itself in the
three-point scattering amplitude of open string momentum modes on the
D2:
\be A_{Non-comm}(p,q,r) (2 \pi)^3\delta^3 (p+q+r) = 
 e^{{i \over 2}  p_1 \theta p_2} A_{Comm}(p,q,r) (2 \pi)^3 \delta^3 (p+q+r) \ .  \ee
From the T-dual, D0-brane perspective, this is a scattering of winding
modes.  The configuration minimizing the world sheet action,
\be S = {1 \over 4 \pi \alpha'} \int g_{ij} dx^i \wedge * dx^j + B_{ij} dx^i \wedge dx^j \label{Bsigma},\ee
can be visualized as a minimal area triangle.
\FIGURE{
\centerline{
\scalebox{.9}{
\begin{picture}(420,258)(0,0)
    \put(0,236){{\bf a)}}
    \put(0,105){{\bf b)}}
    \put(228,236){{\bf c)}}
    \put(18,130){$A$}
    \put(171,130){$B$}
    \put(96,259){$C$}
    \put(292,158){$A$}
    \put(331,71){$B$}
    \put(385,145){$C$}
    \put(277,30){$x_1$}
    \put(230,76){$x_2$}
    \includegraphics{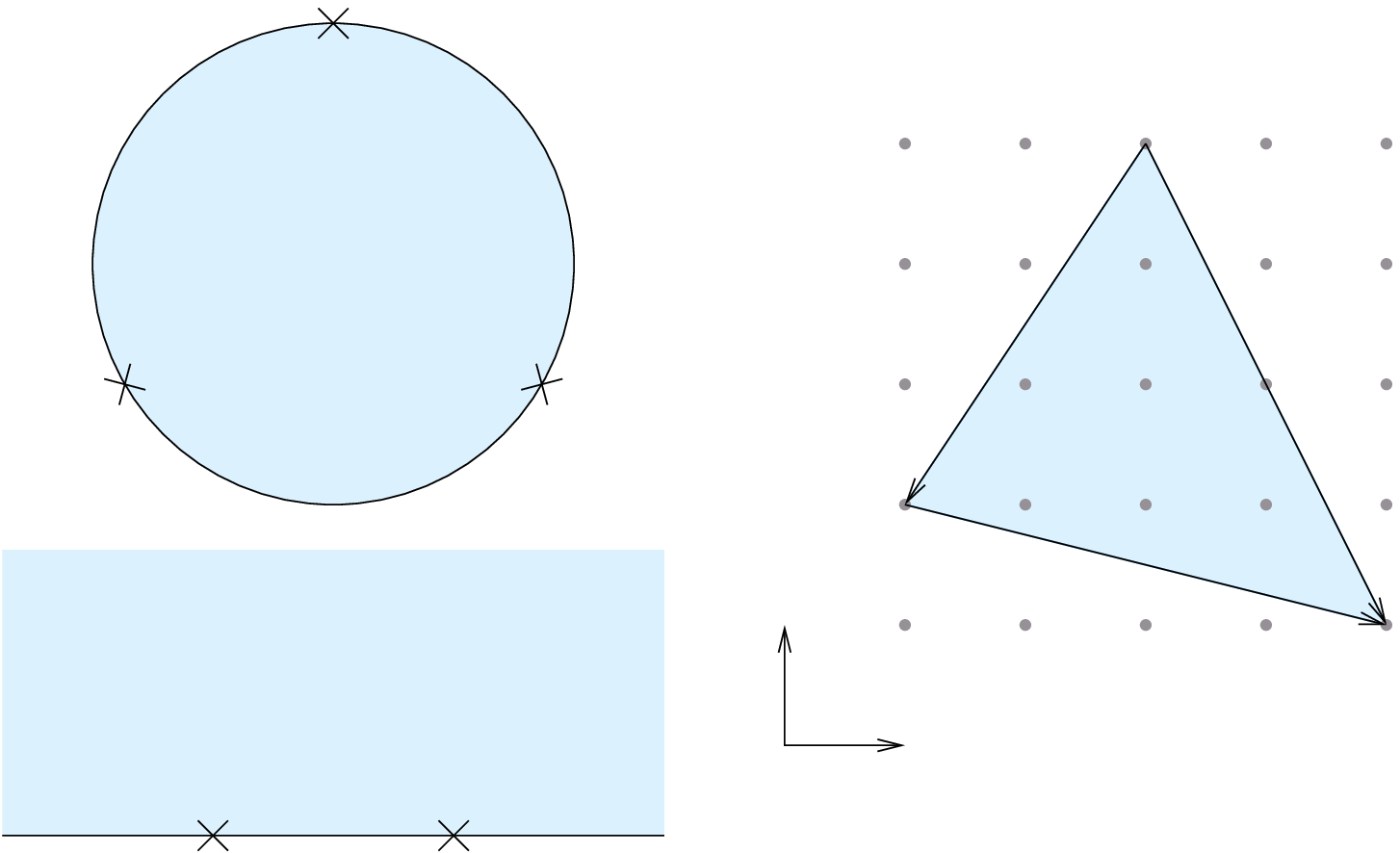}
    \put(-450,73){$\text{UHP}$}
    \put(-402,17){$A$}
    \put(-330,17){$B$}
  \end{picture}
}
}
\caption{\footnotesize{World sheet diagram corresponding to the
scattering of open string winding modes ending on a periodic array of
D-branes. Diagram a) shows the relevant disk geometry with three
punctures representing the open strings.  In b) the same geometry is
shown in the upper half plane.  In c) we show the space-time embedding
of the saddle-point configuration for the constant $B$ background.
The gray circles represent the lattice of D0-branes.  The punctures
$A$, $B$, and $C$ are mapped to the edges of the
triangle.} \label{figa}}
}
When the world sheet is parameterized by the upper half plane, this
triangle corresponds to the configuration,
\be x_i =  {p_i \over 2 \pi i} \log \left({z \over \bar z}\right) +  
 {q_i \over 2 \pi i} \log \left({z -1\over \bar z-1}\right)   
,\qquad  z \in H^+ \label{triangleembedding}
\ee
and is a solution to the equation of motion,
\be \nabla^2 x_i=0 \label{streqm},\ee
which is independent of the $B$-field since $dB = 0$. However, the
world sheet path integral picks up a $B$-dependent phase factor,
\be \exp \left[{i \over 4 \pi \alpha'} \int B_{ij} dx^i \wedge dx^j\right]
= \exp\left[{i \over 2}  p_i  \theta^{ij}q_j \right], \qquad p_i = {2 \pi m_i  \over \tilde L_i}, \qquad q_i = {2 \pi n_i  \over \tilde L_i}, \ee
where
\be \theta = 2 \pi \alpha' B \ .\ee
This phase factor is equivalent to the flux of $B$ through the
triangle illustrated in figure \ref{figa} and allows one to define the
Moyal product,
\be e^{i p x} * e^{i q x} = e^{i p \theta q/2} e^{i (p+q) x} \ . \ee

Non-commutative gauge theory has a clean realization in string theory
as the decoupling limit $\alpha'\to 0$ with $\tilde L_i$ fixed. In
this limit, the masses of open string winding modes remain
finite. Interpreting the winding modes as the momentum modes of the
T-dual picture, one recovers precisely the dynamics of decoupled open
strings in the Seiberg-Witten picture \cite{Seiberg:1999vs}.  When the
background $B$-field is scaled appropriately, the non-commutativity
parameter as seen by the momentum modes in the dual picture also
remains finite.  This is how one reconstructs the effective physics of
non-commutative gauge theories from the winding modes. Of course, in
the case of a constant $B$-field background, one can T-dualize
explicitly and obtain the same non-commutative field theory from
either of the two approaches. In the context of the $\Hspace/\Rspace$
duality, only one approach is available so we use the one to study the
other.

\subsection{Triple T-duality: $\Hspace \leftrightarrow \Rspace$\label{sec3-2}}

Several new feature arise when the $H$-field is non-vanishing. First,
we must consider a three dimensional array of D-branes localized
inside the \Hspace-space. The open string can now wind in three
independent directions.
 
For the non-interacting strings, the mass of the wound strings are
unaffected by the $H$-field. This is because one of the extended
directions of a non-interacting string is the time component, whereas
the $B$-field has no non-vanishing time-like component. Since the
spectrum of the non-interacting wound strings are unaffected by the
$H$-field, one concludes that the geometry of the theory must be
encoded in the interaction terms.

When a string wound along $(m_1, m_2, m_3)$ joins with a
string wound along $(n_1,n_2,n_3)$ to become a string
which winds along $(m_1+n_1, m_2+n_2, m_3+n_3)$, one
expects to find a world sheet configuration which forms a triangle in
a three dimensional lattice, as illustrated in figure \ref{figb}.
\FIGURE{
\centerline{
\begin{picture}(289,181)(0,0)
\put(30,120){$(m_1,m_2,m_3)$}
\put(90,25){$(n_1,n_2,n_3)$}
\put(177,101){$(m_1+n_1,m_2+n_2,m_3+n_3)$}
\put(85,42){$x_2$}
\put(42,86){$x_3$}
\put(17,18){$x_1$}
\includegraphics{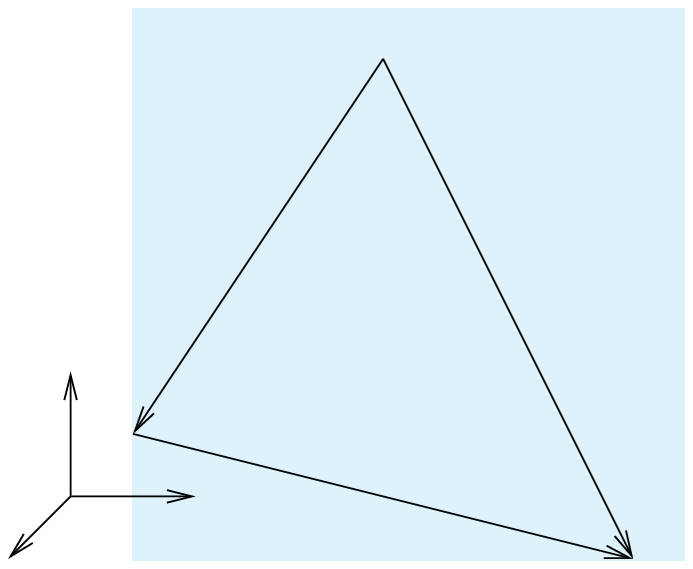}
\end{picture}
}
\caption{\footnotesize{Minimal area triangle configuration for the open string winding modes ending on a three dimensional array of D-branes. The lattice points are suppressed for clarity.\label{figb}}}
}
%
%The lattice points are suppressed for clarity.

If one naively follows the prescription of attributing the flux of
$B$-field through this triangle as a phase, one obtains an expression
for a generalization of the Moyal product. Letting
\be \vec x = (L_1 m_1, L_2 m_2 + L_3 m_3) (1-\sigma_1) + (L_1 n_1 , L_2 n_2, L_3, n_3) \sigma_2 \ee
for $0 < \sigma_1, \sigma_2 < 1$ and $\sigma_1 + \sigma_2 < 1$, one finds\footnote{We adopt a notation similar to the one in \cite{Bouwknegt:2004ap} to facilitate comparison of our algebra with theirs.}
\be \pi(u(m,n)) \equiv e^{{1 \over 4 \pi \alpha'} \int B_{ij} dx^i \wedge dx^j} = 
e^{-2 \pi i N ( {(2m_1 +n_1)/ 6} + c )  (m_2 n_3 - m_3 n_2)} \ . \label{piphase} \ee
The parameter $c$ corresponds to the freedom to move the origin of the
D0-brane lattice along $x_1$, but will not matter in most of the
discussion. One can use this phase to define a new product,

\be e^{i p x} * e^{i q x} = \pi(u(m,n)) e^{i (p + q)x} , \qquad p_i = {2 \pi m_i  \over \tilde L_i}, \qquad q_i = {2 \pi n_i  \over \tilde L_i} \label{naproduct}, \ee
which can also be written in the form,

\be f(x) * g(x) = \left. e^{-{1 \over (2 \pi)^2}  N \tilde L_1 \tilde L_2 \tilde L_3 \left( {2 \partial_{x_1} +\partial_{y_1} \over  6} +C \right)  (\partial_{x_2}\partial_{y_3} - \partial_{x_3}\partial_{y_2})} f(x) g(y) \right|_{x=y} ,\ee
which acts on functions $f(x)$ and $g(x)$ in the
\Rspace-space.\footnote{We are using the fact that points on
\Rspace-space can be viewed as an ordinary $T^3$ when interactions are
ignored.}

An interesting novel feature of this product is that it is
non-associative. Indeed, one can easily confirm that
\be (e^{i p x}* e^{i q x}) * e^{i r x} = \pi (u(m,n)) \pi(u(m+n,l)) e^{i (p+q+r) x} 
\ee
corresponding to the diagram on the left in figure \ref{figc}, and
\be e^{i p x}* (e^{i q x} * e^{i r x}) = \pi (u(m,n+l)) \pi(\alpha_m(u(n,l))) e^{i (p+q+r) x} ,
\ee
corresponding to the diagram on the right in figure \ref{figc}, are not equal since
\be \phi(m,n,l) \equiv {\pi (u(m,n+l)) \pi(\alpha_m u(n,l)) \over \pi (u(m,n)) \pi(u(m+n,l))} =  e^{{i \pi N \over 3} (m \cdot n \times l)} \ne 1 \  . \label{ourphi}\ee
Here, 

\be \alpha_m \pi (u(n,l)) =
e^{-2 \pi i N ( {(2n_1 +l_1)/ 6} +  m/2 + c )  (n_2 l_3 - n_3 l_2)}  \ee
accounts for the fact that $\vec n+\vec l$ does not start at the tail
of $\vec m$.\footnote{If $N$ is even, however, the shift in the phase
is a multiple of $2\pi$ and does not affect the result. The simplicity
of even $N$ is related to the fact that Morita equivalence $\Theta
\rightarrow \Theta + 1$ is an isomorphism which acts non-trivially,
while $\Theta \rightarrow \Theta + 2$ acts trivially.}

The non-associativity can be understood in terms of two different ways of triangulating a quadrilateral, illustrated in figure \ref{figc}.
\FIGURE{
\centerline{
\scalebox{.87}{
\begin{picture}(470,199)(0,0)
\put(75,120){$A$}
\put(115,31){$B$}
\put(224,83){$C$}
\put(156,177){$(A*B)*C$}
\put(300,120){$A$}
\put(340,31){$B$}
\put(449,83){$C$}
\put(381,177){$A*(B*C)$}
\put(9,19){$x_1$}
\put(76,43){$x_2$}
\put(32,85){$x_3$}
\includegraphics{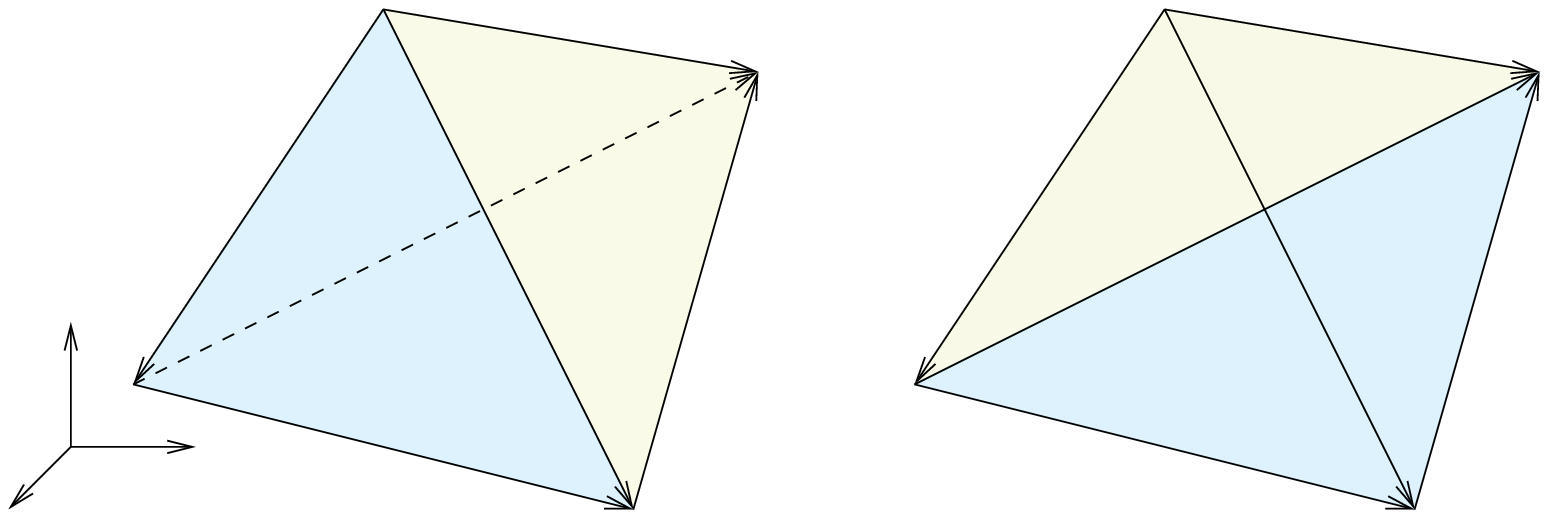}
\end{picture}
}
}
\caption{\footnotesize{Two different triangulations of quadrilateral representing the two different orders of multiplying the open strings $A$, $B$ and $C$. The difference
of $B$-flux through the two triangulations is equivalent to the volume
integral of the $H$-flux through the tetrahedron, and is a measure of
non-associativity of the product (\ref{naproduct}). \label{figc}}}
}
The difference in phases is the surface integral of $B$ over the
tetrahedron, which by Stokes theorem is equivalent to the volume
integral of $H$. This is algebra appears to have the same form as the
twisted crossed product considered in
\cite{Bouwknegt:2004ap}. (Compare with equation (2.3) of
\cite{Bouwknegt:2004ap}). This type of non-associative algebra has
also appeared in other contexts \cite{Jackiw:1984rd} as well as in
attempts to generalize Moyal algebra to higher dimensions in the
context of branes\footnote{We thank Y. Matsuo for telling us about
their on-going work.} \cite{HoMatsuo}.

The picture illustrated in figure \ref{figc} also clarifies the fact
that the non-associative product (\ref{naproduct}) is cyclic, i.e. the
associator is a total derivative. In other words, we have
\be \int d^3 x \, e^{i p x}* (e^{i q x} * e^{i r x}) = 
\int d^3 x \, (e^{i p x}* e^{i q x}) * e^{i r x} = 
(2 \pi)^3 \delta^3(p+q+r) \ . \ee
This is because the conservation of momentum in the \Rspace-space
constrains the tetrahedron to collapse into a triangle.  Cyclicity is
a useful notion in defining field theories using fields whose algebra
is non-associative \cite{Herbst:2003we}. This means that one can
unambiguously write an action whose interaction terms are at most
cubic, such as $\phi^3$ theory.\footnote{A different approach for
defining gauge theories with non-associative fields can be found in
\cite{Ho:2001qk}.}

The non-associative algebra being discussed here comes about somewhat
differently from seemingly similar setup discussed in
\cite{Cornalba:2001sm,Herbst:2001ai}. These authors considered
D3-branes in a non-trivial $H$-field background. Therefore, their
program should be thought of as the study of D3-branes in the
\Hspace-space. What we consider instead is D3-branes in \Rspace
space, or equivalently, D0-branes in \Hspace-space.  The basic setup
is therefore distinct, although certain aspects of the vertex operator
algebra are inevitably similar. It should  also be noted that a D3 in
a presence of an $H$-field acquires an induced magnetic charge
\cite{Pelc:2000kb} via the Hanany-Witten mechanism
\cite{Hanany:1996ie,Witten:1998xy}. Since the \Hspace-space is compact,
additional steps are needed to cancel this induced charge when
constructing a consistent string theory background.

The algebra (\ref{naproduct}) is extremely similar in structure to the
Busby-Smith algebra described in section 3 of
\cite{Bouwknegt:2004ap}. At the present time, it is not completely
clear how one should properly interpret the mathematical formalism
described in \cite{Bouwknegt:2004ap} in physical terms using string
theory.  According to the authors of \cite{Bouwknegt:2004ap}, the
framework described in that paper does not directly concern D-branes
(other than the fact that their charges are encoded by the relevant
K-theory) and should be viewed as a statement regarding the closed
strings.  The fact that we identify similar algebraic structure in the
lattice of dual branes appears to suggest that these structures are
more natural in the context of open strings dynamics.  It is also
worth noting that \cite{Bouwknegt:2004ap} also describes the
\Qspace-space by ``a continuous field of stabilized non-commutative
tori'' that is reminiscent of the description of \Qspace-space in
section \ref{qsec}, which is definitely an open string construction.
It is an important open problem to clarify the proper physical
interpretation of \cite{Bouwknegt:2004ap} and to settle the question
of the relevance of the open v.s.~closed strings.\footnote{This need
not be a mutually exclusive statement since branes can be transmuted
into fluxes and vice versa. Nonetheless, it may turn out that the brane
description is the most natural framework for providing a physical
interpretation of \cite{Bouwknegt:2004ap}.} What is needed is the
analogue of \cite{Connes:1997cr}.

Some of this discussion, however, is purely academic in that the story
of the tetrahedron relied entirely on the triangular world sheet.
Such a world sheet is a solution to the world sheet equation of motion
(\ref{streqm}) when $H=0$, but in the case of non-vanishing $H$-field,
(\ref{streqm}) is corrected to
\be d * d x_i - H_{ijk} dx_j \wedge dx_k = 0 \ . \ee
Interestingly, this equation of motion originally arose in an attempt
to describe vortices in a superfluid \cite{Rasetti:1975ib,Lund:1976ze}
and has resurfaced in various contexts
\cite{Bergshoeff:2000jn,Matsuo:2000fh,Pioline:2002ba,Berman:2004jv}.
The triangular world sheet (\ref{triangleembedding}) is not solved by
this equation. It is not clear if there exist analytic solutions to
this equation, nor is it clear how to use it to properly modify the
naive product described above.

The fact that the non-associative product structure (\ref{naproduct})
is difficult to realize in a fully consistent treatment of critical
string theory may be another indication that \Rspace-space does not
exist as a low-energy effective notion of critical string theory. As a
possible alternative, let us point out that there does exist a
topological sigma model, known as the Poisson-WZ sigma model
\cite{Klimcik:2001vg}, which nicely reproduces our product. The action
of the Poisson-WZ sigma model is given by
\be S = \int \eta_i \wedge dx^i + {2 \pi \over L_1 L_2 L_3} N x_1 dx_2 dx_3, \ee
where $\eta_i$ is a world sheet one-form. Because of the invariance of the action under the gauge transformation,
\be \eta \rightarrow \eta + d \lambda, \ee
one should consider the gauge fixed action,
\be S = \int \eta_i \wedge dx^i  + \eta_i \wedge * d \gamma^i + {2 \pi \over L_1 L_2 L_3} N x_1 dx_2 dx_3 \ . \ee
The fields $\gamma$ and $\eta$ constrain $x$ to be harmonic, making
(\ref{triangleembedding}) the unique solution given the boundary
conditions. The Poisson-WZ model appears to play a role very similar
to the Poisson sigma model \cite{Ikeda:1993fh,Schaller:1994es} whose
boundary correlation functions were shown in \cite{Cattaneo:1999fm} to
elegantly reproduce the deformation quantization formula of Kontsevich
\cite{Kontsevich:1997vb}.  A sophisticated interplay of ideas
involving the Poisson sigma model and the Poisson-WZ model have been
discussed, for example, in
\cite{Park:2000au,Hofman:2002rv,Hofman:2002jz}.  It is quite likely
that the algebraic structure (\ref{naproduct}) and its connection to
\cite{Bouwknegt:2003zg} will turn out to be most transparent in the
context of the open Poisson-WZ sigma model winding modes in a manner
closely resembling the discussion of section \ref{sec3-2}.

Unfortunately, Poisson-WZ sigma model, like the Poisson sigma model,
are too general a construct to embed consistently in critical string
theory.  Not all consistent Kontsevich $*$-deformations are expected
to be realizable as a decoupling limit of a consistent critical
string construction.  The absence of strong conceptual connection
between decoupled open strings and the Poisson sigma model was also
emphasized in \cite{Baulieu:2001fi}.  This is in line with the our
empirical observation that the consistent decoupling limit of
effective field theory on \Rspace-space does not appear to exist.

\section{Concluding Remarks}

In this article, we investigated the possible low-energy effective
description of non-geometric compactifications discussed recently in
the context of novel compactifications
\cite{Hellerman:2002ax,Dabholkar:2002sy,Kachru:2002sk,Shelton:2005cf}. We
followed the guiding principle that generalized notions of geometry
should have a concrete meaning when the intrinsic non-locality
associated with the string scale is decoupled from scale relevant to
geometry.

There are two examples of non-geometric compactifications that arise
naturally in the context of T-duals of $T^3$ with $H$-flux. In the
case of \Qspace-space, we found that the novel non-geometric features
can be understood in terms of a non-commutative geometry compactified
using a Morita duality.  This non-commutative gauge theory has a
singular moduli-space that is resolved by integrating in certain
degrees of freedom. All of these degrees of freedom can be embedded in
a UV complete theory given by little string theory coupled to flavor
matter, but decoupled from gravity.

In the case of the \Rspace-space, we did not find a sensible
decoupling limit.  It is entirely possible that a triple T-duality of
$T^3$ with $H$-flux does not admit a low-energy effective description.
Nonetheless, we did encounter certain features suggestive of
a non-associative generalization of non-commutative geometry.  That
such a generalization to non-commutative geometry should arise in the
\Rspace-space which is seemingly more ``non-geometric'' than the
\Qspace-space is sensible, and it would be extremely
interesting if this connection can be made more precise.

The primary obstacle to understanding the non-associative structure
from critical string theory is the strong gravitational back
reaction. This problem appears to be evaded in the Poisson-WZ sigma
model, which is a topological theory similar to the Poisson sigma
model.  Ultimately, we may find that \Rspace-space has a natural
interpretation in terms of non-associative deformation of space-time
only in the framework the topological sigma model and that this
structure cannot be embedded into critical string theory.

\section*{Acknowledgements}
We would like to thank
T.~Grimm,
N.~Halmagyi,
N.~Itzhaki,
O.~Lunin,
Y.~Matsuo,
J.~Polchinski,
M.~Rozali,
J.~Shelton,
M.~Schulz,
and
W.~Taylor
for discussions, and 
P.~Bouwknegt, 
O.~Ganor,
and
V.~Mathai
for correspondences. This work was supported in part by the DOE grant
DE-FG02-95ER40896 and funds from the University of Wisconsin.

\bibliography{ttwist}

\bibliographystyle{JHEP2}

\end{document}